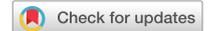

# Higher-order topological semimetal in acoustic crystals

Qiang Wei[1,2,7], Xuewei Zhang[1,2,7], Weiyin Deng [ID][3], Jiuyang Lu [ID][3], Xueqin Huang [ID][3], Mou Yan[3], Gang Chen [ID][1,2,4] ✉, Zhengyou Liu [ID][5,6] ✉ and Suotang Jia [ID][1,2]

The notion of higher-order topological insulators has endowed materials with topological states beyond the first order. Particularly, a three-dimensional (3D) higher-order topological insulator can host topologically protected 1D hinge states, referred to as the second-order topological insulator, or 0D corner states, referred to as the third-order topological insulator. Similarly, a 3D higher-order topological semimetal can be envisaged if it hosts states on the 1D hinges. Here we report the realization of a second-order topological Weyl semimetal in a 3D-printed acoustic crystal, which possesses Weyl points in 3D momentum space, 2D Fermi arc states on surfaces and 1D gapless states on hinges. Like the arc surface states, the hinge states also connect the projections of the Weyl points. Our experimental results evidence the existence of the higher-order topological semimetal, which may pave the way towards innovative acoustic devices.

The discovery of exotic topological states of matter is a thriving research topic in condensed matter physics and material science[1,2]. For a gapped phase, the conventional $d$-dimensional ($d$D) topological insulators feature $(d-1)$D gapless boundary states. Recently, higher-order topological insulators were found to exhibit an extended bulk–boundary correspondence, that is, a $d$D $n$th-order topological insulator has $(d-n)$D boundary states[3–9]. For example, a 2D second-order topological insulator possesses the 0D corner states[3–5], whereas a 3D one hosts the 1D hinge states[6–9]. Although the concept of higher-order topological insulators was first proposed in electronic systems and implemented recently[10], second-order or even first-order topological insulators have been extended and observed in the photonic crystals[11–14], acoustic crystals[15–23] and electric circuits[24,25], benefiting from their macroscopic scale and flexibility of fabrication.

For gapless phases, the topology of the nodal points in 3D momentum space gives rise to the concept of a topological semimetal (TSM)[26], such as the Weyl[27] and Dirac[28] semimetals. Unlike a topological insulator, a conventional 3D TSM is usually characterized by 2D non-closed surface arc boundary states, in contrast to closed surface circle ones. A natural question arises as to whether there exists the 3D higher-order TSM, which hosts the 1D hinge states. Very recently, a few 3D higher-order TSMs were proposed with twofold[5,29,30] or fourfold[31–33] degenerate nodal points, or twofold degenerate nodal loops[33]. However, the higher-order TSMs are yet to be implemented in experiments.

In this work, we report the realization of a 3D second-order TSM (SOTSM) in an acoustic crystal, constructed by stacking a breathing kagomé lattice with double-helix interlayer couplings. The SOTSM hosts the 2D Fermi arc surface states and 1D gapless hinge states, which connect the projections of the Weyl points that result from the $k_z$-dependent polarization protected by the mirror and $C_3$ symmetries. We first illustrate the topological properties of the SOTSM by a tight-binding model, and then present the experimental observation of the Weyl points, the Fermi arc surface states and the hinge states. The theoretical, simulated and experimental results are in good agreement.

We introduce a tight-binding model for the SOTSM. As shown in Fig. 1a, the lattice is constructed by stacking the breathing kagomé lattice along the $z$ direction, in which a unit cell of each layer contains three sites denoted by A (red), B (blue) and C (green). The intralayer couplings contain the intracell hopping $t_a$ (grey) and the intercell hopping $t_b$ (cyan) in the $x$–$y$ plane, whereas the interlayer interaction is dominated by the double-helix hopping $t_z$ (yellow), composed of two equal chiral interlayer couplings that are clockwise and antclockwise. On the basis of sublattices A–C, the Bloch Hamiltonian is written as:

$$H(\mathbf{k}) = \begin{pmatrix} 0 & h_{12} & h_{13} \\ h_{12}^* & 0 & h_{23} \\ h_{13}^* & h_{23}^* & 0 \end{pmatrix} \qquad (1)$$

with $h_{12} = t_a + t_b' e^{-i(k_x/2 + \sqrt{3}k_y/2)a}$, $h_{13} = t_a + t_b' e^{-ik_x a}$ and $h_{23} = t_a + t_b' e^{i(-k_x/2 + \sqrt{3}k_y/2)a}$, where $t_b' = t_b + 2t_z \cos(k_z h)$, $\mathbf{k} = (k_x, k_y, k_z)$ is the Bloch wavevector and $a$ and $h$ are the lattice constants in the $x$–$y$ plane and $z$ direction, respectively. The bandgap closes at $(k_x, k_y) = (\pm 4\pi/3a, 0)$ when $t_a = t_b'$, or $(k_x, k_y) = (0,0)$ when $t_a = -t_b'$. We first discuss the case of $t_a = t_b'$, in which the system has twofold degenerate points at $K_\pm = (4\pi/3a, 0, \pm k_W/h)$ (Fig. 1b) and their time-reversal counterparts $K_\pm' = (-4\pi/3a, 0, \pm k_W/h)$ with $k_W = \arccos[(t_a - t_b)/2t_z]$. As demonstrated in Supplementary Section I, these four degenerate points are the Weyl points with topological charges ±1 (Fig. 1c) and linear dispersions along all three directions.

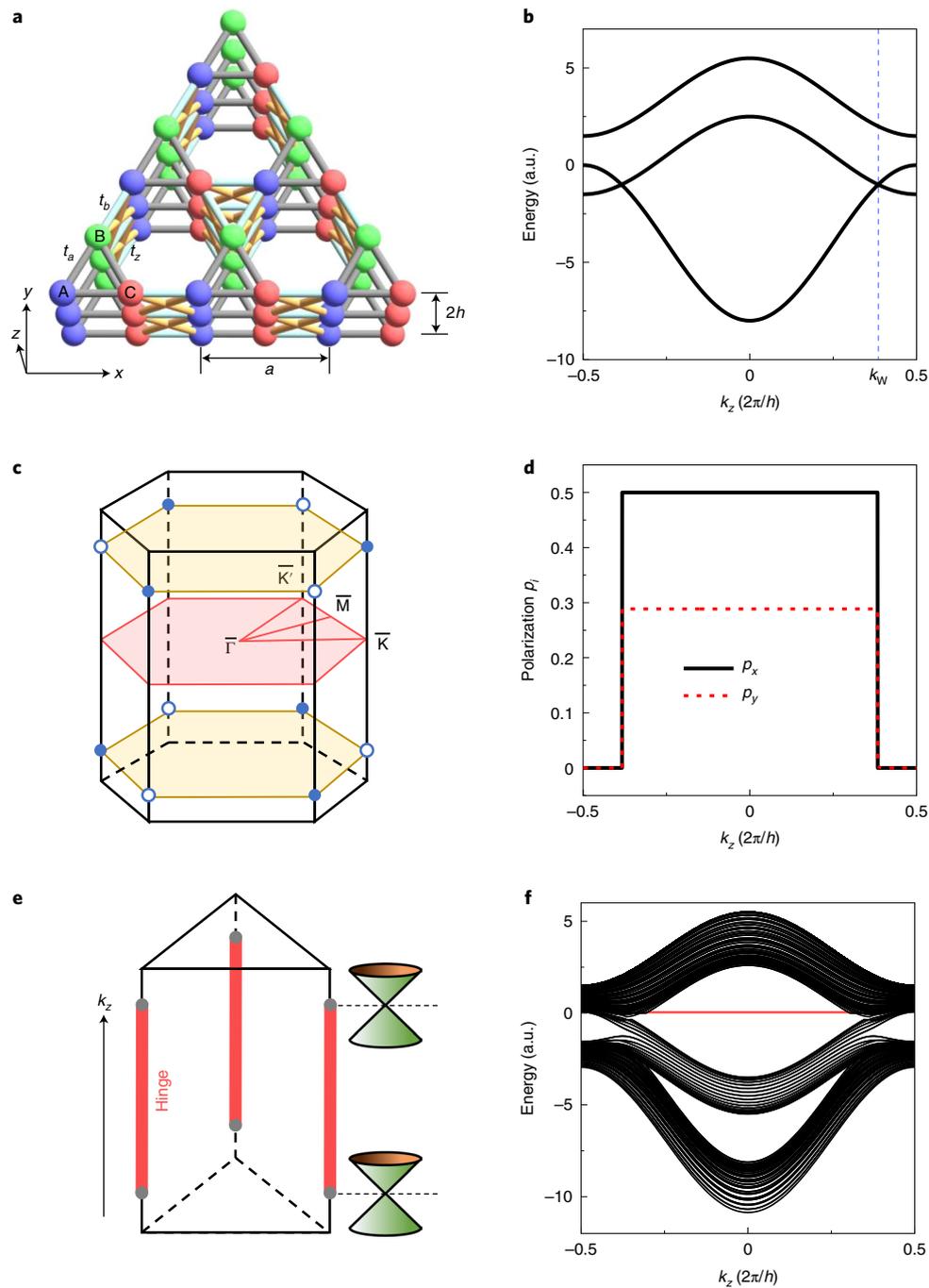

**Fig. 1 | SOTSM for a 3D stacked breathing kagomé lattice. a**, Schematics of the lattice structure with different intralayer ($t_a$ and $t_b$) and interlayer ($t_z$) hoppings. **b**, The bulk state dispersion along the $k_z$ direction with $(k_x, k_y) = (4\pi/3a, 0)$. The dashed blue line shows the position of the degenerate point. **c**, The first Brillouin zone and the distribution of the Weyl points. The hollow and solid circles denote the Weyl points with opposite topological charges. **d**, The polarization $(p_x, p_y)$ of the lowest band along the $k_z$ direction. **e**, Schematics of the hinge states and Weyl points. The vertical direction represents the $k_z$ direction, and the horizontal directions denote real space. **f**, The projected dispersion of a triangle-shaped structure along the $k_z$ direction. The red solid line shows the hinge state dispersion. The parameters in **b**, **d** and **f** were chosen as $t_a = -1$, $t_b = -2.4$ and $t_z = -1$ in arbitrary units (a.u.).

The topological property of our model can be characterized with the 2D topological index by considering $k_z$ as a parameter. The first-order topological index, that is, the $k_z$-dependent Chern number, is zero except for the closing bulk gap at $k_W$. So it is needed to investigate the second-order topological index, the $k_z$-dependent polarization, which is defined as:

$$p_i(k_z) = \frac{1}{S} \iint_{RBZ} A_i d^2k \qquad (2)$$

where $d^2k$ is the area element in the reduced Brillouin zone (RBZ) with area $S$, $A_i = -i\langle u|\partial k_i|u\rangle$ with $i=x,y$, the Berry connection, and $u$ is the Bloch function of the lowest band. The polarization $(p_x, p_y)$ for

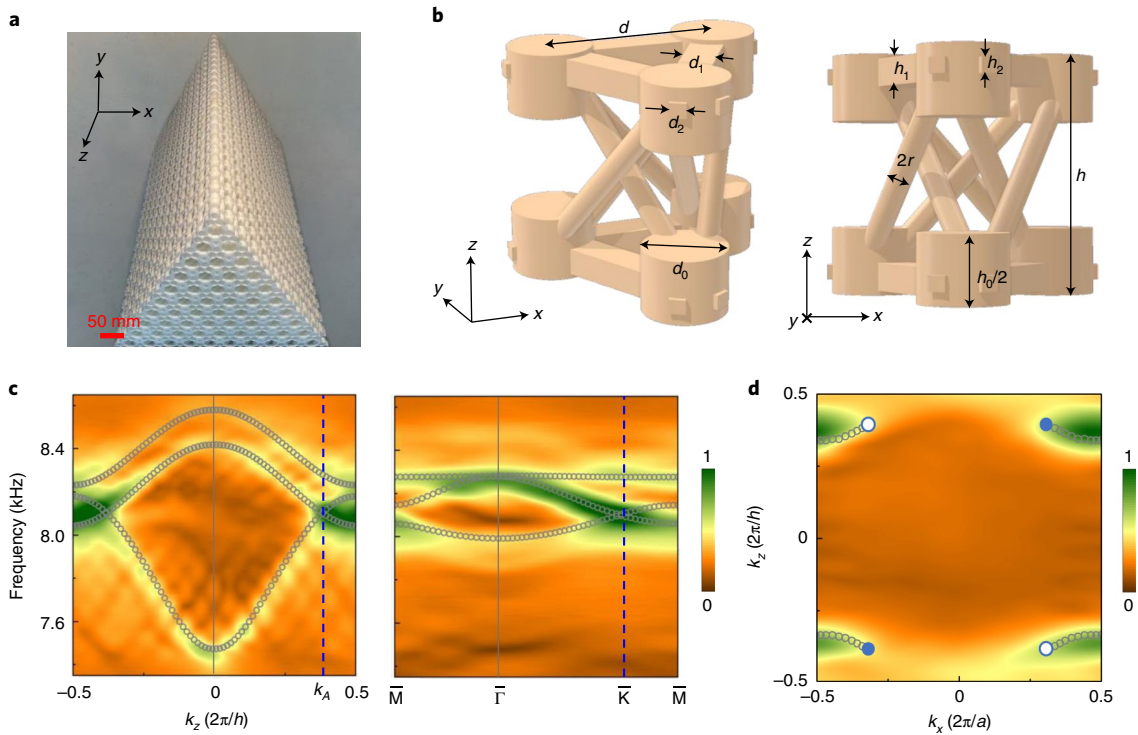

**Fig. 2 | 3D acoustic crystal with Weyl points and Fermi arcs. a**, A photo of the 3D-printed sample. **b**, Schematics of a unit cell, with the side (left panel) and front (right panel) views. **c**, The bulk state dispersions along the $k_z$ direction with $(k_x, k_y) = (4\pi/3a, 0)$ (left panel), and along the high-symmetry lines with $k_z = k_A = 0.38(2\pi/h)$ (right panel). The blue dashed line denotes the position of the Weyl point. **d**, The Fermi arc of the surface states at 8.11 kHz, which connects the projections of the Weyl points (the hollow and solid dots). The colour maps and the grey circles represent the experimental and simulated results, respectively.

a fixed $k_z$ takes a quantized value, because of the mirror and $C_3$ symmetries[5,34]. As shown in Fig. 1d, the polarization is $(1/2, 1/2\sqrt{3})$ for $|k_z| < k_W$ and $(0,0)$ for $|k_z| > k_W$, which corresponds to a phase transition across the Weyl points located at $k_W$. The non-zero polarization gives rise to the hinge states in a triangle-shaped sample with the dispersion connecting the projections of the Weyl points along the $k_z$ direction, as shown by the hinge state distributions and dispersions in Fig. 1e,f. The model unambiguously exhibits the bulk-hinge correspondence and identifies itself a SOTSM.

We implemented the SOTSM in an acoustic crystal. Figure 2a shows a photo of our 3D-printed sample, which is a triangular prism of side 473.65 mm and height 856.68 mm, and comprised 3,465 acoustic cavities inside. Figure 2b shows a unit cell with the lattice constants $a = 44$ mm and $h = 38.9$ mm, which clearly exhibits a double-helix layer-stacking structure. There are three cylindrical cavities of diameter $d_0 = 14$ mm and height $h_0 = 21.5$ mm, separated by a distance $d = 28.6$ mm. The intralayer couplings were introduced by two types of rectangular tubes, whose widths and heights were $d_1 = 5.6$ mm and $h_1 = 4.48$ mm, and $d_2 = 2.7$ mm and $h_2 = 2.43$ mm. The interlayer couplings were induced by double-helix tubes of radius $r = 1.9$ mm. With the cavities viewed as the lattice sites and the connecting tubes as the hoppings, the acoustic crystal can be mapped into the tight-binding model aforementioned.

We demonstrated the Weyl points of the acoustic crystal sample by simulations and experiments. The simulations were performed with the commercial COMSOL Multiphysics solver package, whereas the experimental dispersions were obtained by Fourier transforming the measured acoustic pressure fields of the bulk waves (Methods). The left panel of Fig. 2c shows the dispersions along the $k_z$ direction with $(k_x, k_y) = (4\pi/3a, 0)$. The colour scales and the circles represent the experimental and simulated results, respectively. One can see that there exist two linear crossing points at $k_z = \pm k_A$ with $k_A h/2\pi = 0.38$. In the right panel of Fig. 2c, we show the dispersions along the high-symmetric lines in the $k_x - k_y$ plane with $k_z = k_A$, in which the linear crossing point appears at the $\bar{K}$ point. These results indicate that the crossing point at $(k_x, k_y, k_z) = (4\pi/3a, 0, k_A)$ is the Weyl point with linear dispersions in all three directions, consistent with those of the tight-binding model with the fitting parameters, as discussed in Supplementary Section II. As this system has time-reversal and twofold-rotation (along the $y$ axis) symmetries, the Weyl points are located, respectively, at the $\bar{K}$ and $\bar{K}'$ points and at $k_z = \pm k_A$. This means that this acoustic crystal hosts four Weyl points that reside at the same frequency and is an ideal Weyl semimetal.

It is known that there may exist Fermi arc surface states in the Weyl semimetals[35,36]. Figure 2d shows the Fermi arc surface dispersion at 8.11 kHz. The colour maps represent the measured data and the grey solid lines denote the simulated equifrequency contour, whereas the hollow and solid dots denote the projections of the Weyl points with opposite topological charges. We also simulated the surface states on the $x$–$z$ surface, as shown by a red ellipse in Supplementary Fig. 4a. Note from Supplementary Fig. 4b that the surface states as a whole are gapless. To be more explicit, the surface states cross the projections of the Weyl points and touch the upper and lower bulk bands in the $k_x - k_z$ plane only for $k_z = \pm k_A$. For other $k_z$ values, the surface states do not cross and touch the bulk bands, which is because the $k_z$-dependent Chern number is zero. However, when considering the second-order topological index, we found a non-zero $k_z$-dependent polarization $(p_x, p_y) = (1/2, 1/2\sqrt{3})$ for $|k_z| < k_A$, but zero polarization $(p_x, p_y) = (0,0)$ for $|k_z| > k_A$, as shown in Supplementary Fig. 7a, which indicates that our acoustic crystal sample is a second-order Weyl semimetal. It is informative to further investigate the boundary states on the hinges.

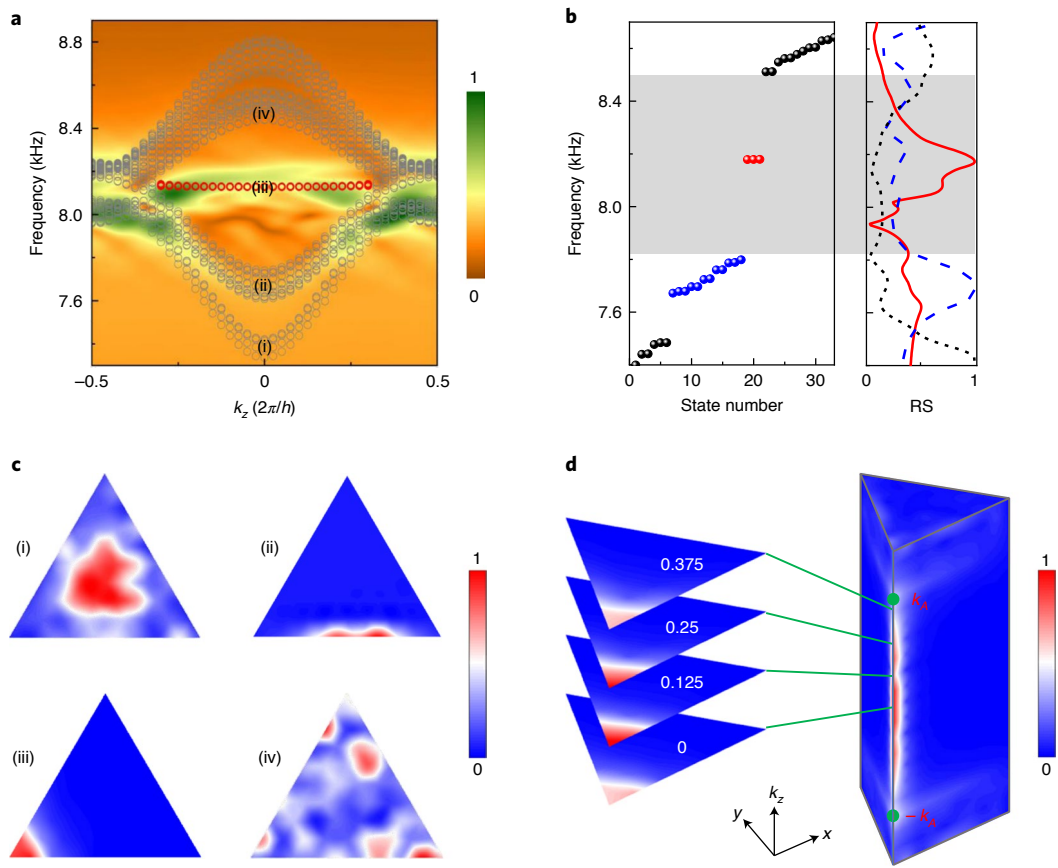

**Fig. 3 | Hinge state and acoustic pressure fields. a**, The projected dispersion along the $k_z$ direction. The red circles represent the simulated hinge state dispersion, and the experimental data are captured by the colour maps. **b**, The simulated eigenfrequencies (left panel) and the measured response spectra (RS) of the acoustic pressure fields (right panel) for $k_z = 0$. The red, blue and black colours denote the hinge, surface and bulk states, respectively. **c**, The measured acoustic pressure fields for the bulk (7.33 kHz and 8.78 kHz), surface (7.70 kHz) and hinge (8.12 kHz) states at $k_z = 0$, which correspond to (i, iv), (ii) and (iii), respectively, in **a**. **d**, Left panel: the measured acoustic pressure fields for $k_z h/2\pi = 0$, 0.125, 0.25 and 0.375 at 8.12 kHz. Right panel: the simulated acoustic pressure fields along the $k_z$ direction at the same frequencies.

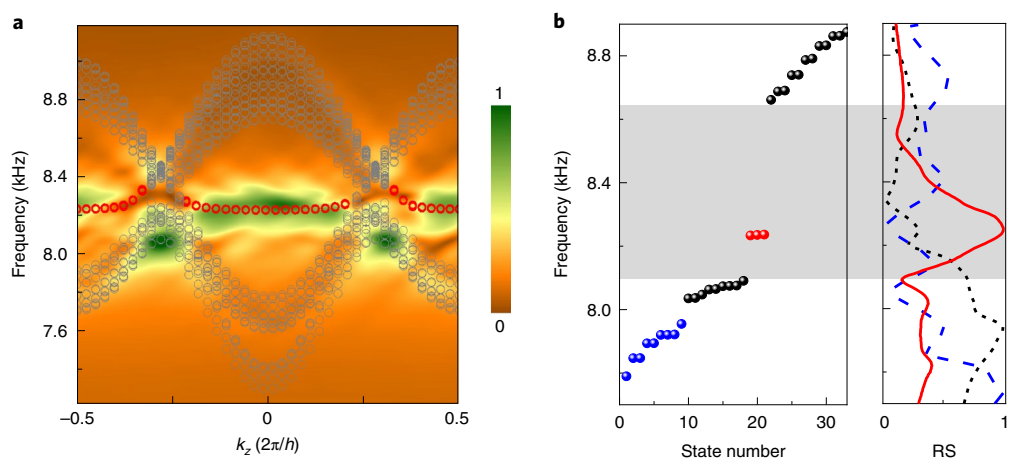

**Fig. 4 | Hinge states and response spectra for different structural parameters. a**, The projected dispersion along the $k_z$ direction. The colour map denotes the experimental data, and the red circles represent the simulated results. **b**, The simulated eigenfrequencies (left panel) and measured response spectra (RS) of the acoustic pressure fields (right panel) for $k_z h/2\pi = 0.5$. The red, blue and black colours denote the hinge, surface and bulk states, respectively.

To excite the hinge states, we placed a headphone in the middle of the hinge of the acoustic crystal sample, and scanned the acoustic field distributions along the hinge with a microphone.

For the experimental set-up and details, refer to and Methods and Supplementary Fig. 8. The projected dispersion along the $k_z$ direction was obtained by Fourier transformation, as shown by the colour map in Fig. 3a. It can be seen that the hinge states exist at a frequency around 8.17 kHz, which agrees well with the simulated ones marked by the red circles (in the range $|k_z| < k_A$). As the excitation was placed at the hinge, the bulk and surface states were poorly stimulated. This can be improved by placing the headphone at the centre of the bulk (to excite the bulk states) or at the surface (to excite the surface states), as shown in Supplementary Fig. 9. Figure 3a, together with Supplementary Fig. 9, gives the full dispersion of all the states, that is, the hinge, surface and bulk states, that existed in our sample. In the left panel of Fig. 3b, we show the simulated eigenstates in our sample for $k_z = 0$. Correspondingly, the measured responses of the bulk, surface and hinge states, with the excitation at the bulk, surface and hinge, respectively, are given in the right panel. It can be seen that the response spectrum (red curve) for the hinge states exhibits a peak at 8.17 kHz, which agrees with that in the simulations (red solid circles).

The existence of the hinge states can be more directly and clearly revealed by the acoustic pressure field distributions. In Fig. 3c, we present the acoustic pressure fields for four different frequencies at $k_z = 0$, which were obtained by extracting the $k_z = 0$ components from the Fourier spectra of the measured field distributions inside the sample for each frequency. For panels (i) and (iv), the exciting headphone was placed in the bulk, whereas for panels (ii) ((iii)) it was placed on the surface (hinge) (Fig. 3c). It can be seen that in panel (ii), the field is localized at the hinge, which indicates the existence of hinge states. Panels (i) and (iv) correspond to the excitations of the bulk states, and panel (ii) corresponds to the excitation of the surface states. To have a complete view of the hinge states, we further show the acoustic pressure fields for varying $k_z$ in Fig. 3d, and the left (right) panel is the experiment (simulation). We observe that the hinge states manifest themselves well for $|k_z| < k_A$. For comparison, we also give the simulated acoustic fields of the bulk and surface states versus $k_z$ in Supplementary Fig. 10.

The full-wave simulations, shown in Fig. 3b and Supplementary Fig. 7b, give three degenerate hinge states, localized respectively at three hinges, which are related by the $C_3$ symmetry. This means that the three hinge states were vanishingly coupled due to the sufficiently large size of the sample. When we put the headphone at a particular hinge (for example, the lower left hinge in the experiment), only the state at this hinge can be stimulated, as shown in panel (ii) of Fig. 3c and Fig. 3d. Otherwise, putting the headphone at any other hinge only excites the state at the corresponding hinge.

Finally, we briefly discuss the case of $t_a = -t'_b$ of the Hamiltonian in equation (1). In this case, besides the Weyl points at $K_\pm$ and $K'_\pm$, the Hamiltonian also hosts the threefold degenerate points at $\Gamma_\pm = (0,0,\pm k_T/h)$, where $k_T = \arccos[-(t_a + t_b)/2t_b] > k_W$. Correspondingly, there appears a new hinge state dispersion that connects the threefold degenerate points, in addition to the first one (for the details, see Supplementary Section VIII). To observe the new hinge states practically, we fabricated a new sample, based on the original one with adjusted structural parameters: $a = 44$ mm, $h = 38.94$ mm, $d_0 = 14$ mm, $h_0 = 21.5$ mm, $d = 28.6$ mm, $d_1 = 3.5$ mm, $h_1 = 3.5$ mm, $d_2 = 3.5$ mm, $h_2 = 3.5$ mm and $r = 3$ mm. Figure 4a shows the measured hinge state dispersions along the $k_z$ direction at a frequency of around 8.23 kHz, which agrees with the simulated ones marked by the red circles. Figure 4b shows the response spectrum of the acoustic pressure field of the new hinge state for $k_z h/2\pi = 0.5$. These response spectra also exhibit a peak at around 8.23 kHz, consistent with the simulated one.

In conclusion, motivated by the pioneering theoretical predictions[5,31], we realized a 3D acoustic SOTSM, which hosts 1D gapless hinge states and exhibits a bulk–hinge correspondence. Our work concretes the higher-order TSMs[5,31,32,37,38], with fundamental significance and potential practical applications. In addition, with the flexibility in obtaining the opposite couplings, the layer-stacking method in our work may be used to realize other types of higher-order TSMs that require both positive and negative hoppings.

## Online content

Any methods, additional references, Nature Research reporting summaries, source data, extended data, supplementary information, acknowledgements, peer review information; details of author contributions and competing interests; and statements of data and code availability are available at https://doi.org/10.1038/s41563-021-00933-4.

## Methods

**Numerical simulations.** All the numerical simulations of an acoustic crystal were performed with the commercial COMSOL Multiphysics solver package. The acoustic crystals were filled with air with a mass density of $1.18\,kg\,m^{-3}$ and sound velocity of $341\,m\,s^{-1}$ at room temperature. Owing to the huge acoustic impedance mismatch compared with air, the 3D-printed plastic material was considered to be a hard boundary.

**Experimental measurements.** A sub-wavelength headphone (diameter 3.0 mm) was placed in the middle of the hinge of the 3D sample for the hinge-state excitations, whereas it was placed at the centre of the corresponding surface or bulk for surface or bulk wave excitations. To measure the acoustic pressure field, a subwavelength microphone (diameter 1.5 mm) attached to the tip of a stainless-steel rod was inserted into the sample and controlled manually. Both the source and receiver were connected to a vector network analyser (Keysight 5061B), where the sound signals (S-parameter S21) were sent and recorded. The network analyser not only generated the excitation signal (a sinusoidal wave sweeps from 4.70 to 11.70 kHz), but also collected the recorded signals with average processing (16 times). The hinge, surface and bulk state dispersions were obtained by Fourier transforming the corresponding measured fields. The response spectra of the hinge, surface and bulk states were obtained by extracting the $k_z = 0$ components from the Fourier spectra of the corresponding measured fields along a line (in the $z$ direction) that passed through the position of the headphone. As only the positions, rather than the heights, of the peaks matter, all the contours (Figs. 2c,d, 3a,c,d and 4a) and the response spectra (Figs. 3b and 4b) were normalized by their maxima.

## Data availability

Owing to their larger size, the data represented in Fig. 3 and Supplementary Fig. 10 are available on Zenodo at https://zenodo.org/record/4441748#.YAFkdznisuV. Source data are provided with this paper.

## Acknowledgements

This work is supported by the National Key R & D Program of China (Grant no. 2017YFA0304203), the National Natural Science Foundation of China (Grant nos 11890701, 11674200, 11704128, 11774275, 11804101, 11974120, 11974005, 12034012, 12074128, and 12074232) and the Shanxi '1331 Project' Key Subjects Construction.

## Author contributions

G.C., Z.L. and S.J. conceived the idea. Q.W. and X.Z. calculated the theoretical results, designed the experiments and carried out the numerical simulations. Q.W., X.Z. and M.Y. performed the experiments. W.D., J.L. and X.H. guided the experimental measurement and analysed the experimental data. G.C. and Z.L. supervised the project. All the authors contributed to the preparation of the manuscript.

## Competing interests

The authors declare no competing interests.

## Additional information

**Supplementary information** The online version contains supplementary material available at https://doi.org/10.1038/s41563-021-00933-4.

**Correspondence and requests for materials** should be addressed to G.C. or Z.L.

**Peer review information** *Nature Materials* thanks Alexander Khanikaev and the other, anonymous, reviewer(s) for their contribution to the peer review of this work.

**Reprints and permissions information** is available at www.nature.com/reprints.

## Supplementary information

# Higher-order topological semimetal in acoustic crystals

# Supplementary Information for

# Higher-order topological semimetal in acoustic crystals

Qiang Wei[1,2*], Xuewei Zhang[1,2*], Weiyin Deng[3], Jiuyang Lu[3], Xueqin Huang[3], Mou Yan[3],

Gang Chen[1,2,4†], Zhengyou Liu[5,6†], Suotang Jia[1,2]

[1]State Key Laboratory of Quantum Optics and Quantum Optics Devices, Institute of Laser spectroscopy, Shanxi University, Taiyuan 030006, China

[2]Collaborative Innovation Center of Extreme Optics, Shanxi University, Taiyuan, Shanxi 030006, China

[3]School of Physics and Optoelectronics, South China University of Technology, Guangzhou, Guangdong 510640, China

[4]Collaborative Innovation Center of Light Manipulations and Applications, Shandong Normal University, Jinan 250358, China

[5]Key Laboratory of Artificial Micro- and Nanostructures of Ministry of Education and School of Physics and Technology, Wuhan University, Wuhan 430072, China

[6]Institute for Advanced Studies, Wuhan University, Wuhan 430072, China

*These authors contributed equally to this work.

†Corresponding author. Email: chengang971@163.com; zyliu@whu.edu.cn

**The Supplementary Information includes:**

**S-I. Weyl points and Fermi arc**

In the main text, we have shown the energy band of the Bloch Hamiltonian in equation (1) possesses the two-fold degenerate points at $K_\pm$ and $K'_\pm$. Here we apply the degenerate perturbation theory to investigate the linear band dispersion and the loop distribution of Berry curvature in momentum space of the low-energy effective Hamiltonian expanding around these degenerate points. It will be demonstrated that these degenerate points are the Weyl points, which generate the so-called Fermi arc on the surface.

We first consider the degenerate point at $K_+$. Assuming $\boldsymbol{k} = \boldsymbol{q} + K_+$, where $\boldsymbol{q} = (q_x, q_y, q_z)$ is the infinitesimal momentum, we construct the perturbation Hamiltonian around $K_+$ as

$$\Delta H = H(\boldsymbol{q} + K_+) - H(K_+). \tag{S1}$$

Hence, the low-energy effective Hamiltonian is given by

$$h_{K_+}(\boldsymbol{q}) = \begin{pmatrix} \langle \psi_1 | \Delta H | \psi_1 \rangle & \langle \psi_1 | \Delta H | \psi_2 \rangle \\ \langle \psi_2 | \Delta H | \psi_1 \rangle & \langle \psi_2 | \Delta H | \psi_2 \rangle \end{pmatrix}, \tag{S2}$$

where $\psi_1$ and $\psi_2$ are the eigenfunctions of the lowest two bands of the Bloch Hamiltonian in equation (1) of the main text. From the Hamiltonian in equation (S2), we plot the bulk state dispersion around $K_+$, as shown in Figs. S1a and S1b. The bulk state dispersion around $K_-$ is similar to the case around $K_+$ and thus not plotted here. Obviously, the bulk state dispersion around the degenerate points along any direction is linear. Furthermore, the Berry curvature is given by

$$F = \frac{\partial A_y}{\partial q_x} - \frac{\partial A_x}{\partial q_y}, \tag{S3}$$

where $A_\mu = -i\langle \phi | \nabla_\mu | \phi \rangle$ is Berry connection, with $\mu = x, y$ and $\phi(\boldsymbol{q})$ being its wavefunction. Figures S1c and S1d show that the flux of the Berry curvature flows from $K_+$ to $K_-$, which is similar to the magnetic monopole in momentum space. So, $K_+$ and $K_-$ are a pair of the Weyl points with opposite charge, denoted by the hollow and solid spheres.

The Weyl points result in Fermi arcs, which are the equi-energy contours of the surface states at a fixed energy. Figure S2a shows the Fermi arcs at the energy of the Weyl points. Because of the ideal property, i.e., all the four Weyl points are at the same energy, the Fermi arcs connect two Weyl points with opposite charges in the absence of coexisting with the bulk states.

In Figs. S2b-S2f, we plot the surface state dispersions along the $k_x$ direction for $k_z h/2\pi = 0$ (b), $0.15$ (c), $0.30$ (d), $0.37$ (e), and $0.45$ (f). It can be found that the surface states as a whole are gapless and touch the upper and lower bulk band at the Weyl points with $k_z h/2\pi = \pm 0.37$. For other $k_z$, the surface states do not cross and touch the bulk bands, which is because that the $k_z$-dependent Chern number is zero. So it is needed to explore the second-order topological index to describe the topological property of this model, as illustrated in the main text.

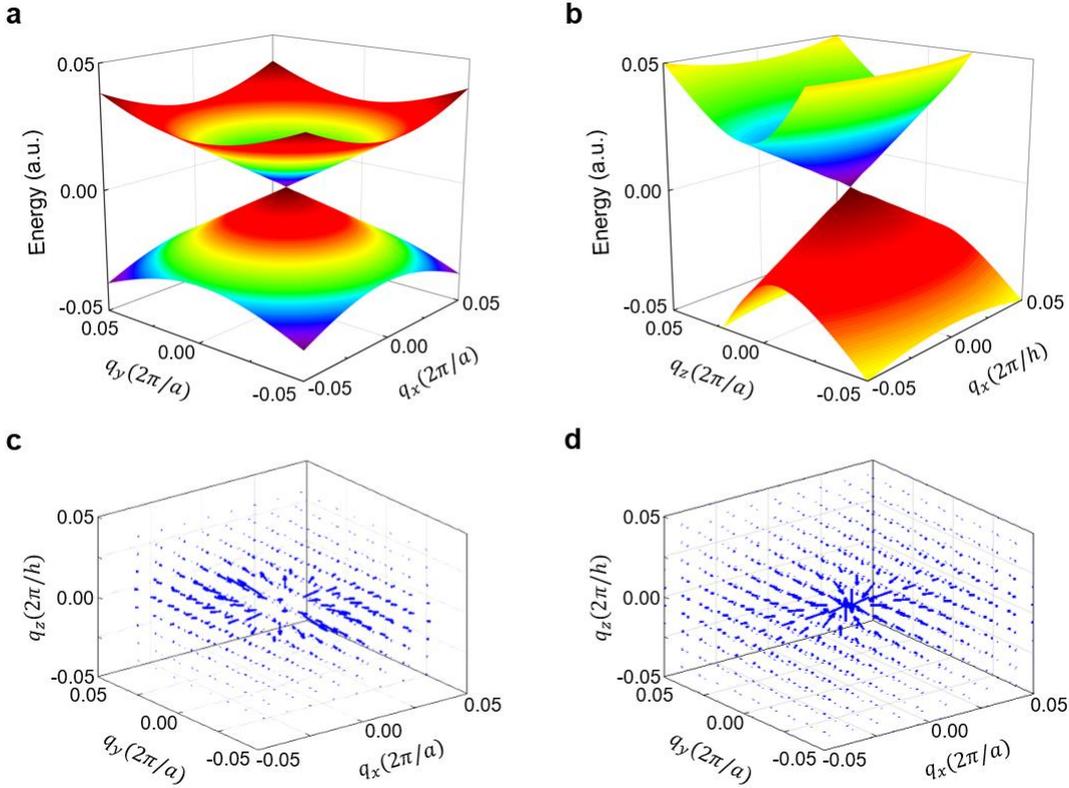

**Figure S1 | Bulk state dispersion and Berry curvature around the Weyl points.** **a-b**, The bulk state dispersions around the Weyl point $K_+$ in the $q_x$-$q_y$ (**a**) and $q_x(q_y)$-$q_z$ (**b**) planes. **c-d**, The spatial distributions of the Berry curvature around $K_+$ (**c**) and $K_-$ (**d**).

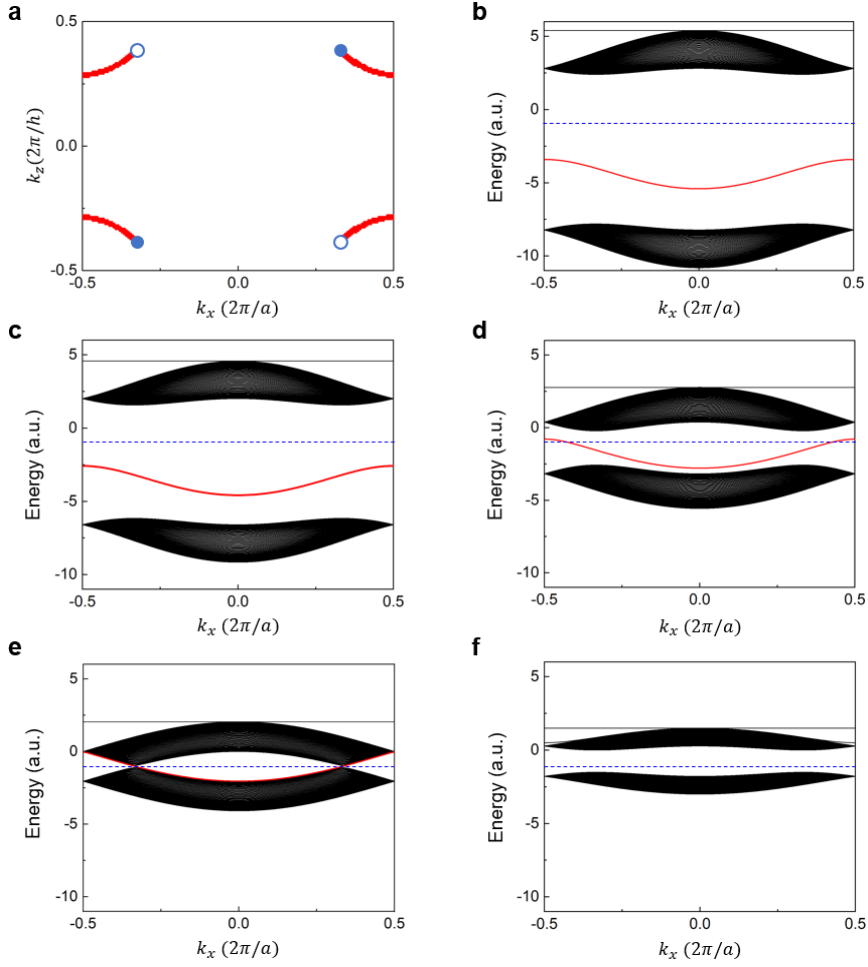

**Figure S2 | Fermi arcs and surface state dispersions. a**, The equi-energy contours of the surface states at the energy of the Weyl points ($E = -1$). The hollow and solid circles denote the Weyl points with opposite topological charges. **b-f**, The surface state dispersions along the $k_x$ direction for $k_z h / 2\pi = 0$ (**b**), 0.15 (**c**), 0.30 (**d**), 0.37 (**e**), and 0.45 (**f**). The red solid lines denote the surface state dispersions and the blue dashed lines show the position of $E = -1$. The plotted parameters are the same as those in <span style="color:blue">Fig. 1</span> of the main text.

## S-II. Fitting parameters near the Weyl point

Based on the tight-binding Hamiltonian in equation (1) of the main text, the bulk Hamiltonian of the acoustic crystal can be described by

$$H_A(\boldsymbol{k}) = f_0 +$$

$$\begin{pmatrix} 0 & t_a + t'_b e^{-i(k_x/2+\sqrt{3}k_y/2)a} & t_a + t'_b e^{-ik_x a} \\ t_a + t'_b e^{i(k_x/2+\sqrt{3}k_y/2)a} & 0 & t_a + t'_b e^{i(-k_x/2+\sqrt{3}k_y/2)a} \\ t_a + t'_b e^{ik_x a} & t_a + t'_b e^{-i(-k_x/2+\sqrt{3}k_y/2)a} & 0 \end{pmatrix},$$

with $t'_b = t_b + 2t_z \cos(k_z h)$, where $a = 44$ mm and $h = 38.9$ mm are the lattice constants in the $x$-$y$ plane and $z$ direction, respectively. The global shift frequency $f_0$ is determined by the frequency of the Weyl point. The corresponding dispersion relations read

$$f_1 = f_0 - t_a - t'_b,$$

$$f_{2,3} = f_0 + (t_a + t'_b)/2 \pm \sqrt{9(t_a^2 + t'^2_b) - 6t_a t'_b + 8t_a t'_b F}/2,$$

with $F = \cos(k_x a) + 2\cos(k_x a/2)\cos(\sqrt{3}k_y a/2)$. Here, we focus on the Weyl points locating at $(4\pi/3a, 0, k_A)$, where $k_A$ fulfills the condition

$$t_a = t'_b = t_b + 2t_z \cos(k_A h). \tag{S4}$$

And the gradients along the $k_x$, $k_y$, and $k_z$ directions are

$$\frac{\partial f_3}{\partial k_x} = \frac{\partial f_3}{\partial k_y} = -\frac{\sqrt{3}}{2}a t_a, \tag{S5}$$

$$\frac{\partial f_3}{\partial k_z} = -2h t_z \sin(k_A h). \tag{S6}$$

In the full-wave simulation, we have $\partial f_3/\partial k_x = \partial f_3/\partial k_y = 41.15$ Hz·m and $\partial f_3/\partial k_z = 184.47$ Hz·m, and $k_A h/2\pi = 0.38$, which lead to $t_a = -47.52$ Hz, $t_b = -166.71$ Hz, $t_z = -158.91$ Hz, and $f_0 = 8.14$ kHz. Based on these fitting parameters, the energy band of the tight-binding model agrees with that obtained by the full-wave simulation, as shown in Fig. S3.

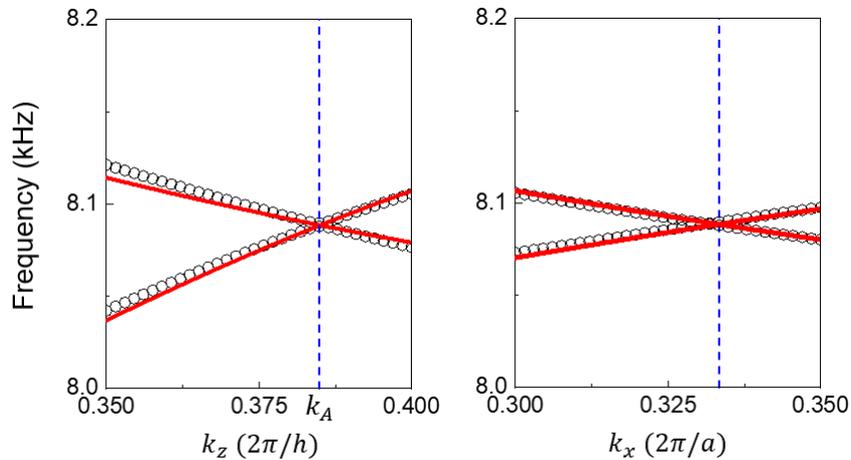

**Figure S3 | The bulk state dispersions near the Weyl point.** Left panel: along the $k_z$ direction. Right panel: along the $k_x$ direction. The red solid lines and black circles represent the analytical results of equations (S4)-(S6) and the full-wave simulations, respectively. The blue dashed lines denote the position of the Weyl point.

## S-III. The simulated surface state dispersions

To calculate the surface state dispersions, we choose a supercell without a top cavity, as shown by Fig. S4a. In Fig. S4b-S4e, the projected dispersions of the surface states in the $k_x$-$k_y$ plane and along the $k_x$ direction for different $k_z$ are plotted. It can be seen that the surface states as a whole are gapless. To be more explicit, the surface states cross the projections of the Weyl points and touch the upper and lower bulk bands on the $k_x$-$k_z$ plane only for $k_z h/2\pi = \pm 0.38$. For other $k_z$, the surface states do not cross and touch the bulk bands, which is because that the $k_z$-dependent Chern number is zero.

In the conventional first-order topological matters, there exists bulk-surface correspondence principle that guarantees the existence of two surface bands in the gap of the bulk state. This should be extended to bulk-hinge correspondence principle for the higher-order topological matters [1, 2]. In such case, the surface state dispersions depend on the choice of the cut of the supercell. Similar to Ref. [3], in our manuscript we choose the supercell without a top cavity as shown in Fig. S4a and the left panel of Fig. S5a, which leads to only one surface band shown in Fig. S4b-S4e and the right panel of Fig. S5a. When another supercell with a top cavity is chosen as in the left panel of Fig. S5b, similar to Ref. [4], two surface bands emerge [4], as shown in the right panel of Fig. S5b. Since the end cuts at bottom of both two choices consist to the boundaries of our finite sample, both the two choices are available.

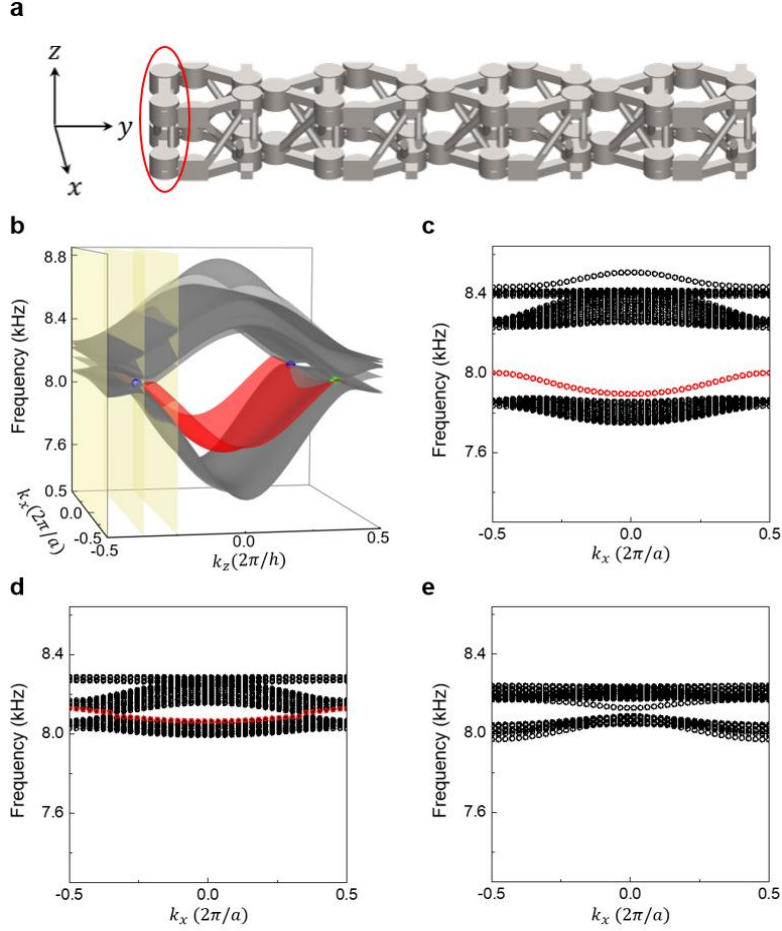

**Figure S4 | The surface state dispersions for the supercell without a top cavity. a**, The supercell. **b**, The projected dispersions in the $k_x$-$k_z$ plane (**b**) and along the $k_x$ direction for $k_z h/2\pi = -0.25$ (**c**), $-0.38$ (**d**), and $-0.5$ (**e**), respectively. The red ellipse in **a** denotes the left boundary, contributing to the surface state dispersions shown by the red region and circles. The blue and green spheres in **b** denote the projected Weyl points, while the yellow planes represent $k_z h/2\pi = -0.25$, $-0.38$, and $-0.5$, respectively.

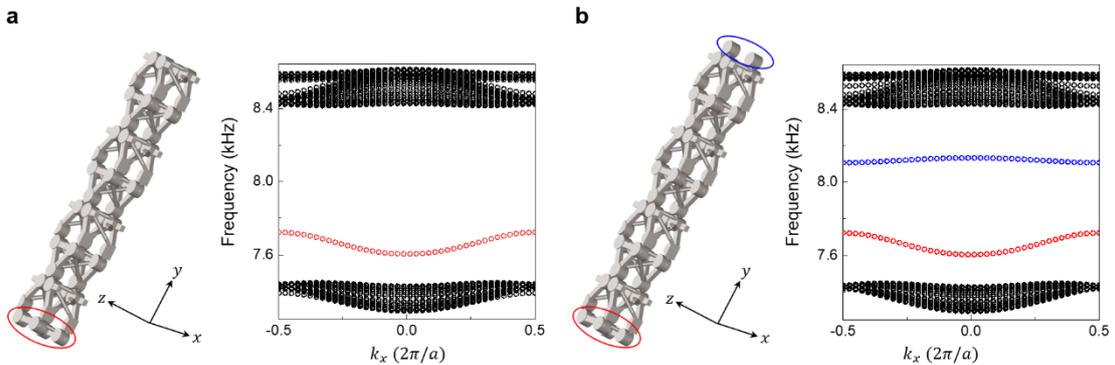

**Figure S5 | The supercells and corresponding projected dispersions for $k_z = 0$. a,** The supercell without a top cavity. **b,** The supercell with a top cavity. The red ellipse denotes the bottom boundary, contributing to the red surface state dispersion, while the blue one represents the top boundary with a top cavity, hosting the blue surface state dispersion.

## S-IV. The simulated bulk polarization

Based on the Wannier band, we calculate the polarization $(p_x, p_y)$ from the COMSOL simulation. We first consider the polarization $p_x$. For a fixed $k_{z0}$, we divide the $k_x$-$k_y$ Brillouin zone into $L_x \times L_y$ parts with $L_x a = N_x dk_x = 2\pi$ and $L_y a = N_y dk_y = 4\sqrt{3}\pi/3$, where $a$ is the lattice constant in the $x$-$y$ plane. Then, the polarization $p_x$ can be written as [3]

$$p_x(k_{z0}) = \frac{1}{S} \int_{\mathrm{BZ}} -i\langle \psi | \partial_{k_x} | \psi \rangle dk_x dk_y$$

$$= \frac{1}{S} \sum_{m=1}^{N_y} dk_y \sum_{n=1}^{N_x} \left\{ \mathrm{Im} \left[ \ln \left( \left\langle \psi_{k_{xn+1}, k_{ym}, k_{z0}} \middle| \psi_{k_{xn}, k_{ym}, k_{z0}} \right\rangle \right) \right] \right\}$$

$$= \frac{1}{N_y} \sum_{m=1}^{N_y} W_x(k_y, k_{z0}),$$

where

$$W_x(k_y, k_{z0}) = \frac{1}{L_x a} \sum_{n=1}^{N_x} \left\{ \mathrm{Im} \left[ \ln \left( \left\langle \psi_{k_{xn+1}, k_{ym}, k_{z0}} \middle| \psi_{k_{xn}, k_{ym}, k_{z0}} \right\rangle \right) \right] \right\}.$$

The polarization $p_y$ can be calculated by the same way. For simplicity, we take $k_{z0} h/2\pi = 0$ or $0.5$ with $N_x = N_y = 20$ as two examples, where $h$ is the lattice constant in the $z$ direction. For our acoustic crystal, the lattice constants $a = 44$ mm and $h = 39.9$ mm. Figure S6 shows the calculated Wannier bands $W_x(k_y, 0)$, $W_x(k_y, \pi/h)$, $W_y(k_x, 0)$, and $W_y(k_x, \pi/h)$. Finally, for the obtained Wannier bands of each $k_z$, the polarization $(p_x, p_y) = (1/2, 1/2\sqrt{3})$ for $|k_z| < k_A$ and $(p_x, p_y) = (0,0)$ for $|k_z| > k_A$, as shown in Fig. S7a. This nontrivial polarization indicates that there exists three hinge states, as shown in Fig. S7b.

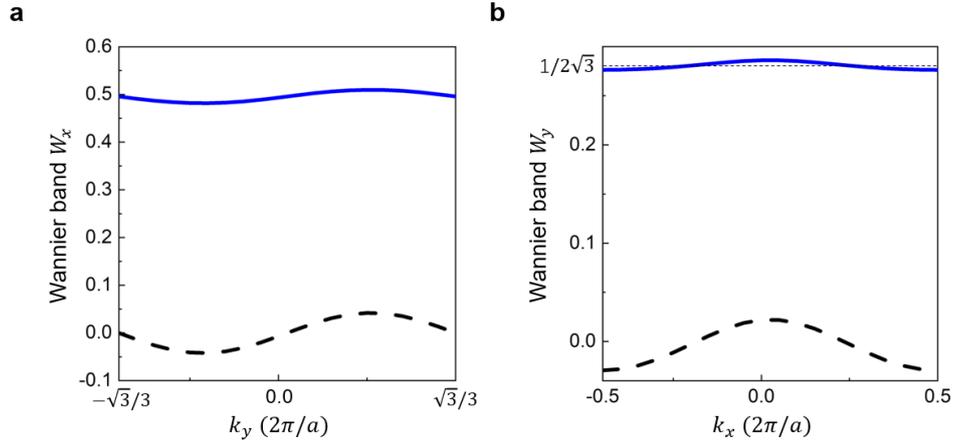

**Figure S6 | The simulated Wannier bands. a**, and **b**, $W_x(k_y)$ and $W_y(k_x)$ for $k_z h/2\pi = 0$ (blue solid line) and $0.5$ (black dashed line), respectively.

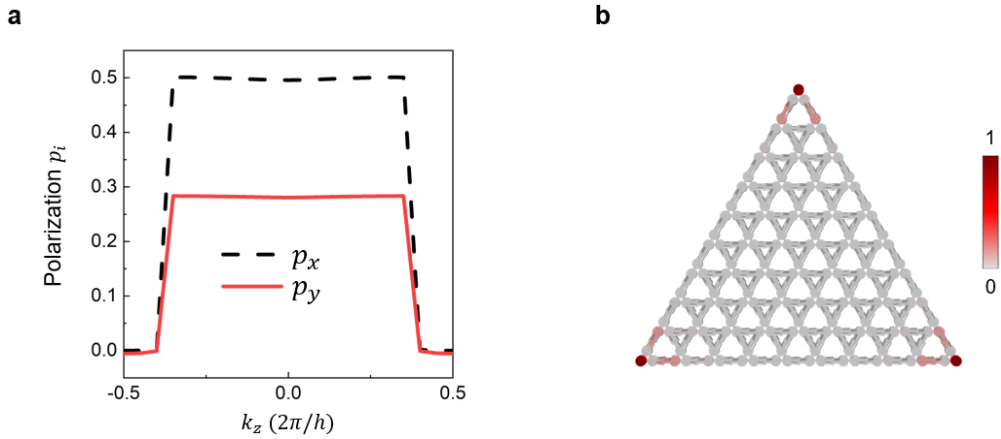

**Figure S7 | a**, The simulated polarization $(p_x, p_y)$ along the $k_z$ direction. **b**, The simulated local density of state at the frequency $8.17$ kHz.

## S-V. The schematic experimental setup to obtain the hinge state dispersions

The measurement scheme is provided in Fig. S8. To excite the hinge band, the point source is placed at the middle of the hinge. The receiver is inserted into the cavity of the hinge through the side tube to measure the acoustic field distributions of the hinge. When the acoustic pressure fields are collected by the network analyzer, the hinge state dispersions are obtained by preforming the Fourier transformation.

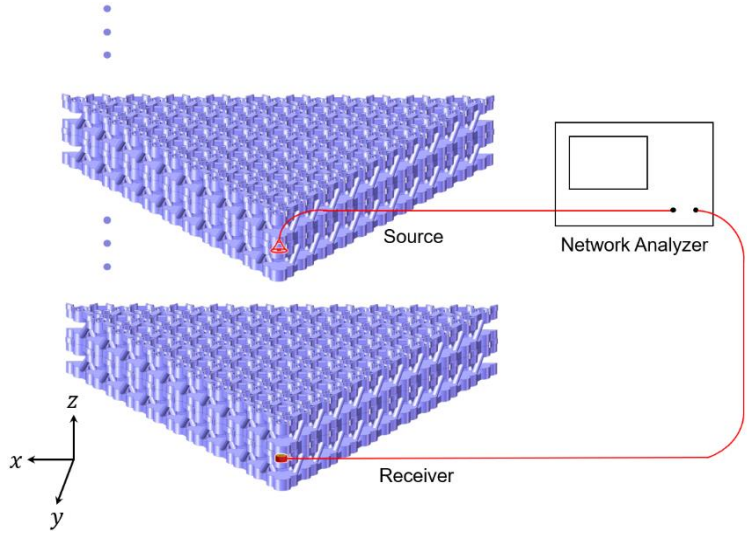

**Figure S8 | Schematic experimental setup.** The point source is placed at the middle of the hinge to excite the hinge band. The receiver is inserted into the cavity of the hinge through the side tube to measure the acoustic field distributions. The network analyzer is used to collect the amplitude and phase of the acoustic pressure fields.

## S-VI. The measured projected dispersions of the bulk and surface states

To measure the projected dispersions of the bulk and surface states, a source is placed in the center of the surface and bulk. The experimental observations are shown in Fig. S9.

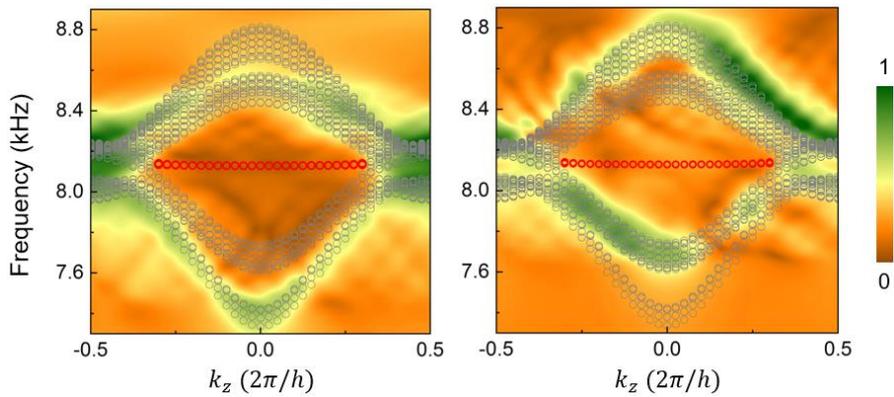

**Figure S9 | The measured projected dispersions along the $k_z$ direction.** Left panel: The bulk state. Right panel: The surface state. The color maps and gray circles denote the experimental and simulated results, respectively.

## S-VII. The 3D field distributions of the bulk and surface states

Figure S10 shows the 3D simulated acoustic pressure field distributions of the bulk and surface states.

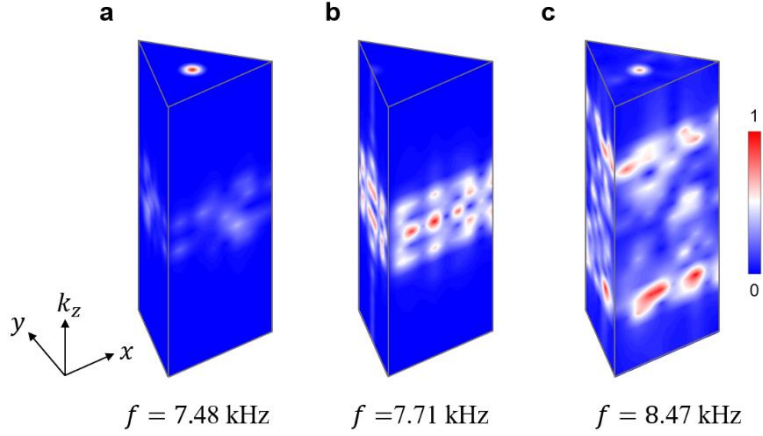

**Figure S10 | The 3D plot of the simulated acoustic pressure field distributions**. **a** and **c**, The bulk states. **b**, The surface state.

## S-VIII. The SOTSM for $t_a = -t_b'$

As shown in the main text, we have observed the hinge states between the Weyl points for $t_a = t_b'$. We now consider another case of $t_a = -t_b'$, in which the SOTSM hosts both the four Weyl and two three-fold degenerate points. The Weyl points are still located at $K_\pm = (4\pi/3a, 0, \pm k_W/h)$ and their time-reversal counterparts $K_\pm'$ with $k_W = \arccos[(t_a - t_b)/2t_z]$, while the three-fold degenerate points occur at $\Gamma_\pm = (0, 0, \pm k_T/h)$ with $k_T = \arccos[-(t_a + t_b)/2t_z] > k_W$. The distributions of the Weyl and three-fold degenerate points in the first Brillouin zone are shown in Fig. S11. In Fig. S12a, we show the bulk state dispersion along the $k_z$ direction at $(k_x, k_y) = (4\pi/3a, 0)$, verifying the existences of the Weyl points crossing between the lower two bands at $k_z = \pm k_W$. And Fig. S12b plots the bulk state dispersion along the $k_z$ direction at $(k_x, k_y) = (0, 0)$, where two of the three bands are degenerate (blue curve), showing the three-fold degenerate points at $k_z = \pm k_T$. Similar to that of $t_a = t_b'$, the first-order topology for this case is also trivial. In Fig.

S12c, we calculate the bulk polarization is $\left(1/2, 1/2\sqrt{3}\right)$ for $|k_z| < k_W/h$ and $k_T/h < |k_z| \leq \pi/h$, and $(0,0)$ for $k_W/h < |k_z| < k_T/h$. According to bulk-hinge correspondence principle [1-2], the connection of the three-fold degenerate points generates new hinge states, apart from the ones between the Weyl points, as shown in Fig. S12d.

To observe the new hinge states practically, we fabricate a new sample, based on the original one, with adjusted structural parameters: $a = 44$ mm, $h = 38.94$ mm, $d_0 = 14$ mm, $h_0 = 21.5$ mm, $d = 28.6$ mm, $d_1 = 3.5$ mm, $h_1 = 3.5$ mm, $d_2 = 3.5$ mm, $h_2 = 3.5$ mm, and $r = 3$ mm. Based on these parameters, we calculate the bulk band structure (Figs. S13a and S13b) and the projected dispersion along the $k_z$ direction (Fig. S13c) by the full-wave simulation. It can be seen clearly that this acoustic crystal sample exhibits new hinge states connected by two three-fold degenerate points located at $(k_x, k_y, k_z) = \left(0, 0, \pm 0.35(2\pi/h)\right)$. As shown by the main text, these hinge states are confirmed by measuring their dispersions and the response spectra of the acoustic pressure fields for $k_z h/2\pi = 0.5$.

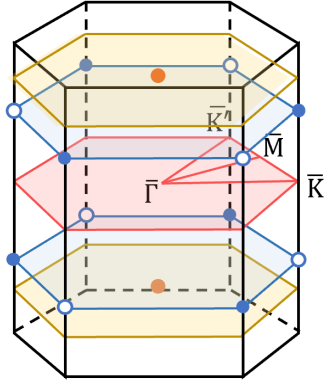

**Figure S11 | The distributions of the Weyl and three-fold degenerate points in the first Brillouin zone.**

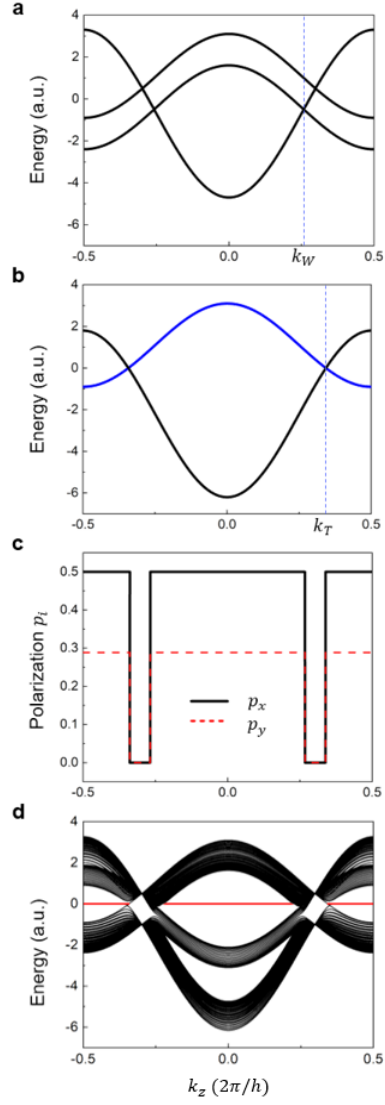

**Figure S12 | The bulk state dispersions and polarization, and the hinge state dispersions. a**, The bulk state dispersion along the $k_z$ direction with $(k_x, k_y) = (4\pi/3a, 0)$. The blue dashed line shows the position $(k_z = k_W)$ of the Weyl point of the lower two bands. **b**, The bulk state dispersion along the $k_z$ direction with $(k_x, k_y) = (0,0)$. The blue solid and dashed line shows the two-fold degenerate band and the position $(k_z = k_T)$ of the three-fold degenerate point. **c**, The polarization $(p_x, p_y)$ of the lowest band along the $k_z$ direction. **d**, The projected dispersion of a triangle-shaped sample along the $k_z$ direction. The red solid lines denote the hinge state dispersions. The parameters are chosen as $t_a = -0.5$, $t_b = -0.6$, and $t_z = -1$.

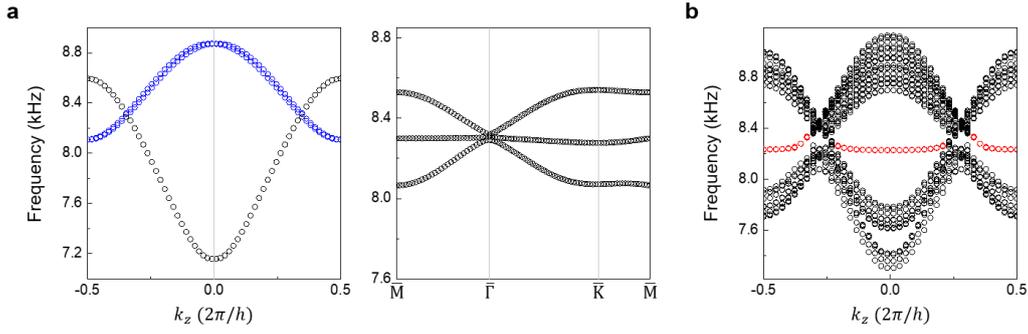

**Figure S13 | The bulk and hinge state dispersions of the acoustic crystal sample.**
**a**, The bulk state dispersion along the $k_z$ direction with $(k_x, k_y) = (0,0)$ (left panel) and along the high-symmetry lines with $k_z h/2\pi = \pm 0.35$ (right panel). The blue circles denote the two-fold degenerate band. **b**, The projected dispersion along the $k_z$ direction. The red circles represent the hinge state dispersions.